\documentstyle[12pt]{article}
\begin{document}

\title {THE WEYL TENSOR AND EQUILIBRIUM CONFIGURATIONS OF SELF--GRAVITATING FLUIDS}
\author{Luis Herrera\thanks{Postal address: Apartado 80793, Caracas 1080A,
Venezuela; E-mail address: laherrera@telcel.net.ve}\\
Escuela de F\'{\i}sica. Facultad de Ciencias.\\ Universidad Central de Venezuela. Caracas, Venezuela.\\
}
\date{}
\maketitle

\begin{abstract}
It is shown that (except for two well defined cases), the necessary and sufficient condition for any spherically symmetric
distribution of fluid to leave  the state of equilibrium (or quasi--equilibrium), is that the Weyl tensor changes with respect to its 
value in the state of equilibrium (or quasi--equilibrium).
\end{abstract}
\newpage
\section{Introduction}
Since the publication of Penrose'work \cite{penrose}, there has been an increasing interest in the possible role  of Weyl tensor
(or some function of it) in the evolution of self-gravitating systems (\cite{weyl} and references therein). This interest is 
reinforced by the fact that (at least) for spherically symmetric distributions of fluid, the Weyl tensor may be expressed exclusively in 
terms of the density contrast and the local anisotropy of the pressure \cite{inho}, which in turn are known to affect the fate of
gravitational collapse \cite{anis}.

In this work we shall consider spherically symmetric fluid configurations which are initially in equilibrium (or quasi--equilibrium),
and we shall look for necessary and sufficient conditions to leave such regime. Usually, these conditions are expressed through
the adiabatic index, (or some function of it) which indicates the variation of pressure with density, for a given fluid element
( see \cite{adiabatic} and references therein).

As we shall see here, such departure from equilibrium (or quasi--equilibrium) is allowed if and only if, the Weyl 
tensor within the fluid distribution , changes with respect to its value in the initial (equilibrium or quasi--equilibrium) state.

There is however two possible exceptions for this. One is represented by the ``inflationary'' equation of state $\rho+p=0$. 
The other situation when our result does not apply, corresponds to the case when then system leaving the equilibrium enters into dissipative regime and $\kappa T/\tau(\rho+p)=1$,
 where $\kappa$,$T$ and $\tau$ denote respectively the thermal conduction, the temperature and the thermal relaxation time. 
Both situations imply from the ``dynamical'' point of view, that the effective inertial mass of any fluid element vanish
(see \cite{eim} and references therein).

The above mentioned result is reminiscent , in some sense, of a one recently obtained ,
 regarding the conditions for a transition from a non-dissipative to a dissipative regime in a FRW-flat model\cite{FRW}. 
Indeed in that case, the Weyl tensor is always zero, and such transition is only allowed if $\kappa T/\tau(\rho+p)=1$.

The manuscript is organized as follows.In the next section we give the expressions for the field and transport equations, and for the Weyl tensor. In Section III the conditions for 
the departure from equilibrium (and quasiequilibrium) are found.
Finally in the last Section the results are discussed.

\section{Field and Transport equations}
\subsection{Field equations}
\noindent
We consider spherically symmetric distributions of collapsing 
fluid, which for sake of completeness we assume to be anisotropic, 
undergoing dissipation in the form of heat flow, bounded by a 
spherical surface $\Sigma$.

\noindent
The line element is given in Schwarzschild-like coordinates by
 
\begin{equation}
ds^2=e^{\nu} dt^2 - e^{\lambda} dr^2 - 
r^2 \left( d\theta^2 + sin^2\theta d\phi^2 \right)
\label{metric}
\end{equation}

\noindent
where $\nu(t,r)$ and $\lambda(t,r)$ are functions of their arguments. We 
number the coordinates: $x^0=t; \, x^1=r; \, x^2=\theta; \, x^3=\phi$.

\noindent
The metric (\ref{metric}) has to satisfy Einstein field equations
 
\begin{equation}
G^\nu_\mu=-8\pi T^\nu_\mu
\label{Efeq}
\end{equation}

\noindent 
which in our case read \cite{BO}:

\begin{equation}
-8\pi T^0_0=-\frac{1}{r^2}+e^{-\lambda} 
\left(\frac{1}{r^2}-\frac{\lambda'}{r} \right)
\label{feq00}
\end{equation}

\begin{equation}
-8\pi T^1_1=-\frac{1}{r^2}+e^{-\lambda}
\left(\frac{1}{r^2}+\frac{\nu'}{r}\right)
\label{feq11}
\end{equation}

\begin{eqnarray}
-8\pi T^2_2  =  -  8\pi T^3_3 = & - &\frac{e^{-\nu}}{4}\left(2\ddot\lambda+
\dot\lambda(\dot\lambda-\dot\nu)\right) \nonumber \\
& + & \frac{e^{-\lambda}}{4}
\left(2\nu''+\nu'^2 - 
\lambda'\nu' + 2\frac{\nu' - \lambda'}{r}\right)
\label{feq2233}
\end{eqnarray}

\begin{equation}
-8\pi T_{01}=-\frac{\dot\lambda}{r}
\label{feq01}
\end{equation}

\noindent
where dots and primes stand for partial differentiation with respect
to t and r  
respectively.

\noindent
In order to give physical significance to the $T^{\mu}_{\nu}$ components 
we apply the Bondi approach \cite{BO}.

\noindent
Thus, following Bondi, let us introduce purely locally Minkowski 
coordinates ($\tau, x, y, z$)

$$d\tau=e^{\nu/2}dt\,\qquad\,dx=e^{\lambda/2}dr\,\qquad\,
dy=rd\theta\,\qquad\, dz=rsin\theta d\phi$$

\noindent
Then, denoting the Minkowski components of the energy tensor by a bar, 
we have

$$\bar T^0_0=T^0_0\,\qquad\,
\bar T^1_1=T^1_1\,\qquad\,\bar T^2_2=T^2_2\,\qquad\,
\bar T^3_3=T^3_3\,\qquad\,\bar T_{01}=e^{-(\nu+\lambda)/2}T_{01}$$

\noindent
Next, we suppose that when viewed by an observer moving relative to these 
coordinates with velocity $\omega$ in the radial direction, the physical 
content  of space consists of an anisotropic fluid of energy density $\rho$, 
radial pressure $P_r$, tangential pressure $P_\bot$ and radial heat flux 
$\hat q$. Thus, when viewed by this moving observer the covariant tensor in 
Minkowski coordinates is

\[ \left(\begin{array}{cccc}
\rho    &  -\hat q  &   0     &   0    \\
-\hat q &  P_r      &   0     &   0    \\
0       &   0       & P_\bot  &   0    \\
0       &   0       &   0     &   P_\bot  
\end{array} \right) \]

\noindent
Then a Lorentz transformation readily shows that

\begin{equation}
T^0_0=\bar T^0_0= \frac{\rho + P_r \omega^2 }{1 - \omega^2} + 
\frac{2 Q \omega e^{\lambda/2}}{(1 - \omega^2)^{1/2}}
\label{T00}
\end{equation}

\begin{equation}
T^1_1=\bar T^1_1=-\frac{ P_r + \rho \omega^2}{1 - \omega^2} - 
\frac{2 Q \omega e^{\lambda/2}}{(1 - \omega^2)^{1/2}}
\label{T11}
\end{equation}

\begin{equation}
T^2_2=T^3_3=\bar T^2_2=\bar T^3_3=-P_\bot
\label{T2233}
\end{equation}

\begin{equation}
T_{01}=e^{(\nu + \lambda)/2} \bar T_{01}=
-\frac{(\rho + P_r) \omega e^{(\nu + \lambda)/2}}{1 - \omega^2} - 
\frac{Q e^{\nu/2} e^{\lambda}}{(1 - \omega^2)^{1/2}} (1 + \omega^2)
\label{T01}
\end{equation}

\noindent
with

\begin{equation}
Q \equiv \frac{\hat q e^{-\lambda/2}}{(1 - \omega^2)^{1/2}}
\label{defq}
\end{equation}

\noindent
Note that the velocity in the ($t,r,\theta,\phi$) system, $dr/dt$, 
is related to $\omega$ by

\begin{equation}
\omega=\frac{dr}{dt}\,e^{(\lambda-\nu)/2}
\label{omega}
\end{equation}

\noindent
At the outside of the fluid distribution, the spacetime is that of Vaidya, 
given by

\begin{equation}
ds^2= \left(1-\frac{2M(u)}{\cal R}\right) du^2 + 2dud{\cal R} - 
{\cal R}^2 \left(d\theta^2 + sin^2\theta d\phi^2 \right)
\label{Vaidya}
\end{equation}

\noindent
where $u$ is a time-like coordinate such that $u=constant$ is (asymptotically) a 
null cone open to the future and $\cal R$ is a null coordinate ($g_{\cal R\cal R}=0$). It should 
be remarked, however, that strictly speaking, the radiation can be considered 
in radial free streaming only at radial infinity.

\noindent
The two coordinate systems ($t,r,\theta,\phi$) and ($u,\cal R,\theta,\phi$) are 
related at the boundary surface and outside it by

\begin{equation}
u=t-r-2M\,ln \left(\frac{r}{2M}-1\right)
\label{u}
\end{equation}

\begin{equation}
{\cal R}=r
\label{R}
\end{equation}

\noindent
In order to match smoothly the two metrics above on the boundary surface 
$r=r_\Sigma(t)$, we have to require the continuity of the first fundamental 
form across that surface. As result of this matching we obtain

\begin{equation}
\left[P_r\right]_\Sigma=\left[Q\,e^{\lambda/2}\left(1-\omega^2\right)^
{1/2}\right]_\Sigma = \left[\hat q\right]_\Sigma
\label{PQ}
\end{equation}

\noindent
expressing the discontinuity of the radial pressure in the presence 
of heat flow, which is a well known result \cite{Santos}.

\noindent
Next, it will be useful to calculate the radial components of the 
conservation law

\begin{equation}
T^\mu_{\nu;\mu}=0
\label{dTmn}
\end{equation}

\noindent
After tedious but simple calculations we get

\begin{equation}
\left(-8\pi T^1_1\right)'=\frac{16\pi}{r} \left(T^1_1-T^2_2\right) 
+ 4\pi \nu' \left(T^1_1-T^0_0\right) + 
\frac{e^{-\nu}}{r} \left(\ddot\lambda + \frac{\dot\lambda^2}{2}
- \frac{\dot\lambda \dot\nu}{2}\right)
\label{T1p}
\end{equation}

\noindent
which in the static case becomes

\begin{equation}
P'_r=-\frac{\nu'}{2}\left(\rho+P_r\right)+
\frac{2\left(P_\bot-P_r\right)}{r}
\label{Prp}
\end{equation}

\noindent
representing the generalization of the Tolman-Oppenheimer-Volkof equation 
for anisotropic fluids \cite{BL}.

\subsection{Transport equations}

\noindent
As it is well known, the Maxwell-Fourier law for the radiation 
flux, usually assumed in the study of stars interiors, leads to a parabolic equation (diffusion equation) which predicts 
propagation of perturbation with infinite speed (see \cite{JP}--\cite{M} and 
references therein). This simple fact is at the origin of the pathologies 
\cite{HL} found in the approaches of Eckart \cite{E} and Landau \cite{LL} 
for relativistic dissipative processes.

\noindent
To overcome such difficulties, different relativistic 
theories with non-vanishing relaxation times have been proposed 
in the past \cite{IS}--\cite{C}. The important point is that all these 
theories provide a heat transport equation which is not of 
Maxwell-Fourier type but of Cattaneo type \cite{Cattaneo}, leading thereby to a 
hyperbolic equation for the propagation of thermal perturbation.

\noindent
Accordingly we shall describe the heat transport by means of a 
relativistic Israel-Stewart equation \cite{IS} , which reads 

\begin{equation}
\tau \frac{Dq^\alpha}{Ds} + q^\alpha = 
\kappa P^{\alpha \beta} \left(T_{,\beta} - T a_\beta\right) - 
\tau u^\alpha q_\beta a^\beta-
\frac{1}{2} \kappa T^2 
\left(\frac{\tau}{\kappa T^2} u^\beta\right)_{;\beta} q^\alpha
\label{Catrel}
\end{equation}

\noindent
where $\kappa$, $\tau$, $T$, $q^\beta$ and $a^\beta$ denote thermal conductivity, 
thermal relaxation time, temperature, the heat flow vector and the components of the four 
acceleration, respectively. Also, $P^{\alpha \beta}$ is the projector 
onto the hypersurface orthogonal to the four velocity $u^\alpha$.

\noindent
In our case this equation has only two independent components, 
which read, for $\alpha=0$

\[
\tau e^{(\lambda-\nu)/2}
\left(
Q \dot\omega + \dot Q \omega + Q \omega \dot\lambda
\right) +
\tau \left(
Q' \omega^2 + Q \omega \omega' + \frac{Q \omega^2 \lambda'}{2} 
\right)  
\]
\[
+ \frac{\tau Q  \omega^2}{r} 
+ Q \omega e^{\lambda/2} \left(1 - \omega^2\right)^{1/2}  =   
- \, \frac{\kappa \omega^2 \dot T e^{-\nu/2}}
{\left(1 - \omega^2\right)^{1/2}}
- \, \frac{\kappa \omega T' e^{-\lambda/2}}
{\left(1 - \omega^2\right)^{1/2}} 
\]
\[
- \, \frac{\nu'}{2} 
\frac{\kappa T \omega e^{-\lambda/2}}{\left(1 - \omega^2\right)^{1/2}}
- \frac12 Q \omega \left(e^{(\lambda-\nu)/2} \dot{\tau}+\omega \tau'\right)
\]
\[
-\frac12 \tau Q \omega \left[e^{(\lambda-\nu)/2}
\left(\frac{\omega \dot{\omega}}{1-\omega^2}+\frac{\dot{\lambda}}{2}\right)
+\left(\frac{\omega'}{1-\omega^2}+\frac{\nu'\omega}{2}\right)\right]
\]
\[
+\frac12 \tau Q \omega \left[\frac1\kappa
\left(e^{(\lambda-\nu)/2}\dot{\kappa}+\omega\kappa'\right)
+\frac2T\left(e^{(\lambda-\nu)/2}\dot{T}+\omega T'\right)
\right]
\]
\begin{eqnarray}
 + \left(\tau Q e^{(\lambda-\nu)/2} - \,  
\frac{\kappa T \omega e^{-\nu/2}}{\left(1 - \omega^2\right)^{1/2}}\right)
& \times &  
\left(\frac{\omega \dot\lambda}{2} + \frac{\dot\omega}{1 - \omega^2}\right) 
\nonumber \\ 
 + \left(\tau Q - \, 
\frac{\kappa T \omega e^{-\lambda/2}}{\left(1 - \omega^2\right)^{1/2}}\right)
& \times &
 \frac{\omega \omega'}{1 - \omega^2}
\label{com0}
\end{eqnarray}

\noindent
and for $\alpha=1$

\[
\tau e^{(\lambda-\nu)/2}
\left(
\dot Q + \frac{Q \dot\lambda}{2} + \frac{Q \omega^2 \dot\lambda}{2}
\right) + \tau \omega \left(
Q' + \frac{Q \lambda'}{2} \right) 
\]
\[
+\frac{\tau Q \omega}{r} + Q e^{\lambda/2} \left(1 - \omega^2\right)^{1/2} =
- \, \frac{\kappa \omega \dot T e^{-\nu/2}}
{\left(1 - \omega^2\right)^{1/2}}
- \, \frac{\kappa T' e^{-\lambda/2}}
{\left(1 - \omega^2\right)^{1/2}} 
\]
\[
-\, \frac{\nu'}{2}
\frac{\kappa T e^{-\lambda/2}}{\left(1 - \omega^2\right)^{1/2}}
-\, \frac{1}{2} Q \left(e^{(\lambda-\nu)/2}\dot{\tau}+\omega\tau'\right)
\]
\[
-\frac12 \tau Q \left[e^{(\lambda-\nu)/2}
\left(\frac{\omega \dot{\omega}}{1-\omega^2}+\frac{\dot{\lambda}}{2}\right)
+\left(\frac{\omega'}{1-\omega^2}+\frac{\nu'\omega}{2}\right)\right]
\]
\[
+\frac12 \tau Q \left[\frac1\kappa
\left(e^{(\lambda-\nu)/2}\dot{\kappa}+\omega\kappa'\right)
+\frac2T\left(e^{(\lambda-\nu)/2}\dot{T}+\omega T'\right)
\right]
\]
\begin{eqnarray}
 + \left(\tau Q \omega e^{(\lambda-\nu)/2} - \,  
\frac{\kappa T e^{-\nu/2}}{\left(1 - \omega^2\right)^{1/2}}\right)
& \times &  
\left(\frac{\omega \dot\lambda}{2} + \frac{\dot\omega}{1 - \omega^2}\right) 
\nonumber \\ 
 + \left(\tau Q \omega - \, 
\frac{\kappa T e^{-\lambda/2}}{\left(1 - \omega^2\right)^{1/2}}\right)
& \times &
 \frac{\omega \omega'}{1 - \omega^2}
\label{com1}
\end{eqnarray}

\noindent
where the expressions

\begin{equation}
u^\mu=\left(\frac{e^{-\nu/2}}{\left(1-\omega^2\right)^{1/2}},\,
\frac{\omega\, e^{-\lambda/2}}{\left(1-\omega^2\right)^{1/2}},\,0,\,0\right)
\label{umu}
\end{equation}

\begin{equation}
q^\mu=Q\,\left(\omega\,e^{(\lambda-\nu)/2},\,1,\,0,\,0\right)
\label{qmu}
\end{equation}

\noindent
have been used.

\subsection{The Weyl tensor}.
For the next section we shall need the 
components of the Weyl tensor. Using Maple V, it is found
 that all non-vanishing components are 
proportional to 

\begin{eqnarray}
W \equiv \frac{r}{2} C^{3}_{232} & = & W_{(s)} + \frac{r^3 e^{-\nu}}{12} 
\left(\ddot\lambda + \frac{\dot\lambda^2}{2} - 
\frac{\dot\lambda \dot\nu}{2}\right)
\label{W}
\end{eqnarray}

\noindent

where

\begin{equation}
W_{(s)} = 
\frac{r^3 e^{-\lambda}}{6}
\left( \frac{e^\lambda}{r^2} - \frac{1}{r^2} +
\frac{\nu' \lambda'}{4} - \frac{\nu'^2}{4} -
\frac{\nu''}{2} - \frac{\lambda'}{2r} + \frac{\nu'}{2r} \right)
\label{Ws}
\end{equation}

\noindent
 corresponds 
to the contribution in the static (and quasi--static)  case .
Also, the following expression relating the Weyl tensor through the source terms, may be found \cite{inho}
\begin{equation}
W = - \frac{4 \pi}{3} \int^r_0{r^3 \left(T^0_0\right)' dr} + 
\frac{4 \pi}{3} r^3 \left(T^2_2 - T^1_1\right)
\label{Wint}
\end{equation}
\section{Leaving the equilibrium(quasiequilibrium)}
\noindent
Let us now consider a spherically symmetric fluid distribution which 
initially may be in either hydrostatic and thermal equilibrium (i.e. 
$\omega = Q = 0$), or slowly evolving and dissipating energy through 
a radial heat flow vector.

\noindent
Before proceeding further with the treatment of our problem, let us 
clearly specify the meaning of ``slowly evolving''. That means that 
our sphere changes on a time scale which is very large as compared to 
the typical time in which it reacts on a slight perturbation of 
hydrostatic equilibrium. This typical time is called hydrostatic 
time scale. Thus a slowly evolving system is always in hydrostatic 
equilibrium (very close to), and its evolution may be regarded as 
a sequence of static models linked by (\ref{feq01}).

\noindent
As we mentioned before, this assumption is very sensible, since 
the hydrostatic time scale is usually very small.

\noindent
Thus, it is of the order of $27$ minutes for the sun, $4.5$ seconds 
for a white dwarf and $10^{-4}$ seconds for a neutron star of one 
solar mass and $10$ Km radius \cite{adiabatic}.

\noindent
In terms of $\omega$ and metric functions, slow evolution means 
that the radial velocity $\omega$ measured by the Minkowski observer, 
as well as metric time derivatives are so small that their products and 
second order time derivatives may be neglected (an invariant 
characterization of slow evolution may be found in \cite{HS}).

\noindent
Thus \cite{HS}

\begin{equation}
\ddot\nu\approx\ddot\lambda\approx\dot\lambda \dot\nu\approx
\dot\lambda^2\approx\dot\nu^2\approx
\omega^2\approx\dot\omega=0
\label{neg}
\end{equation}

\noindent
As it follows from (\ref{feq01}) and (\ref{T01}), $Q$ is of the 
order $O(\omega)$. 
Thus in the slowly evolving regime, relaxation terms may be neglected 
and (\ref{Catrel}) becomes the usual Landau-Eckart transport equation .

\noindent
Then, using (\ref{neg}) and (\ref{T1p}) we obtain (\ref{Prp}), 
which as mentioned before is the equation of hydrostatic equilibrium 
for an anisotropic fluid. This is in agreement with what was mentioned 
above, in the sense that a slowly evolving system is in hydrostatic 
equilibrium.

\noindent
Let us now return to our problem. Before perturbation, the two 
possible initial states of our system are characterized by:

\begin{enumerate}
\item Static
\begin{equation}
\dot \omega = \dot Q = \omega = Q = 0 
\label{eqdt}
\end{equation}
\item Slowly evolving
\begin{equation}
\dot \omega = \dot Q = 0
\label{evlen}
\end{equation}
\begin{equation}
Q \approx O(\omega) \not = 0 \; \qquad (small)
\label{Qorom}
\end{equation}
\end{enumerate}

\noindent
where the meaning of ``small'' is given by (\ref{neg}).

\noindent
Let us now assume that our system is submitted to perturbations 
which force it to depart from hydrostatic equilibrium but keeping the 
spherical symmetry.

\noindent
We shall study the perturbed system on a time scale which is 
small as compared to the thermal adjustment time.

\noindent
Then, immediately after perturbation (``immediately'' understood 
in the sense above), we have for the first initial condition 
(static) 

\begin{equation}
\omega = Q = 0
\label{omyQ0}
\end{equation}

\begin{equation}
\dot\omega \approx \dot Q \not = 0 \; \qquad (small)
\label{chiq}
\end{equation}

\noindent
whereas for the second initial condition (slowly evolving)

\begin{equation}
Q \approx O(\omega) \not = 0 \; \qquad (small)
\label{Qseg}
\end{equation}

\begin{equation}
\dot Q \approx \dot\omega \not = 0 \; \qquad (small)
\label{pomQ2}
\end{equation}

\noindent
As we shall see below, both initial conditions lead to the same final 
equations.

\noindent
Let us now write explicitly eq.(\ref{T1p}). With the help of 
(\ref{T00})--(\ref{T01}), we find after long but trivial calculations 

\begin{eqnarray}
& & \frac{P_r'}{1-\omega^2} + 
\frac{\rho' \omega^2}{1-\omega^2} + 
\frac{2 \omega \omega' \rho}{1-\omega^2} + 
\frac{2 \omega \omega' P_r}{\left(1-\omega^2\right)^2}   
\nonumber \\ 
& + &
\frac{2 \omega^3 \omega' \rho}{\left(1-\omega^2\right)^2}  +
\frac{2 Q' \omega e^{\lambda/2}}{\left(1-\omega^2\right)^{1/2}} +
\frac{2 Q \omega' e^{\lambda/2}}{\left(1-\omega^2\right)^{1/2}} + 
\frac{2 Q \omega^2 \omega' e^{\lambda/2}}{\left(1-\omega^2\right)^{3/2}}  
\nonumber \\
& + &
\frac{2}{r} \, [ \, 
\frac{4 \pi r^3}{r-2m} \, 
\left(\rho + P_r \omega^2\right) \, 
\frac{Q \omega e^{\lambda/2}}{\left(1-\omega^2\right)^{3/2}} 
+ \frac{12 \pi r^3}{r-2m} \, 
\left(\frac{Q \omega e^{\lambda/2}}{\left(1-\omega^2\right)^{1/2}}\right)^2   
\nonumber \\
& + & 
\left(\rho + P_r\right) \, \frac{\omega^2}{1-\omega^2} 
+ \left(P_r - P_\bot\right)  + 
\frac{2 Q \omega e^{\lambda/2}}{\left(1-\omega^2\right)^{1/2}} +  
\frac{\left(\rho+P_r\right)}{2} \, 
\frac{1+\omega^2}{1-\omega^2} \, \frac{m}{r-2m} 
\nonumber \\ 
& + &
\frac{Q \omega e^{\lambda/2}}{\left(1-\omega^2\right)^{1/2}} \, 
\frac{m}{r-2m} + 
\frac{2 \pi r^3}{r-2m} \, 
\left(P_r+\rho\omega^2\right) \left(\rho+P_r\right) \, 
\frac{1+\omega^2}{\left(1-\omega^2\right)^2} 
\nonumber \\ 
& + &
\frac{8 \pi r^3}{r-2m} \, 
\left(P_r+\rho\omega^2\right) \, 
\frac{ Q \omega e^{\lambda/2}}{\left(1-\omega^2\right)^{3/2}}  + 
\frac{4 \pi r^3}{r-2m} \, 
Q \omega e^{\lambda/2} \left(\rho+P_r\right) \, 
\frac{1+\omega^2}{\left(1-\omega^2\right)^{3/2}} \, ] 
\nonumber \\
& = &
\frac{e^{-\nu}}{8 \pi r} 
\left(\ddot\lambda + 
\frac{\dot\lambda^2}{2} - 
\frac{\dot\lambda \dot\nu}{2}\right)
\label{horror}
\end{eqnarray}

\noindent
which, when evaluated immediately after perturbation, reduces to

\begin{equation}
P'_r + \frac{\left(\rho + P_r\right) m}{r^2 \left(1 - 2m/r\right)} 
+ \frac{4 \pi r}{\left(1 - 2m/r\right)} \left(P_r \rho + P_r^2\right) 
+ \frac{2 \left(P_r - P_\bot\right)}{r} 
= \frac{e^{- \nu}}{8 \pi r} \ddot \lambda
\label{menho}
\end{equation}

\noindent
for both initial states.

\noindent
On the other hand, an expression for $\ddot\lambda$ may be obtained by 
taking the time derivative of (\ref{feq01})

\begin{eqnarray}
\ddot\lambda & = & -  8 \pi r e^{(\nu + \lambda)/2} 
[\,
\left(\rho + P_r\right) \frac{\omega}{1-\omega^2} 
\frac{\dot\nu}{2} + 
Q e^{\lambda/2} \frac{1+\omega^2}{\left(1-\omega^2\right)^{1/2}} 
\frac{\dot\nu}{2} \nonumber \\
& + & 
\frac{\left(\rho+P_r\right) \omega}{1-\omega^2} 
\frac{\dot\lambda}{2} +  
Q e^{\lambda/2} \frac{1+\omega^2}{\left(1-\omega^2\right)^{1/2}} 
\dot\lambda + 
\left(\dot\rho + \dot P_r\right) 
\frac{\omega}{1-\omega^2} \nonumber \\
& + &
\left(\rho+P_r\right) \dot\omega 
\frac{1+\omega^2}{\left(1-\omega^2\right)^{2}}  + 
\dot Q e^{\lambda/2} \frac{1+\omega^2}{\left(1-\omega^2\right)^{1/2}} 
\nonumber \\ 
& + &
Q e^{\lambda/2} \frac{\omega \dot\omega \left(3-\omega^2\right)}
{\left(1-\omega^2\right)^{3/2}}
\,]
\label{pplex}
\end{eqnarray}

\noindent
which, in its turn, when evaluated after perturbation, reads

\begin{equation}
\ddot\lambda = - 8 \pi r e^{(\nu+\lambda)/2} 
\left[\left(\rho+P_r\right) \dot\omega + 
\dot Q e^{\lambda/2}\right]
\label{ddl12}
\end{equation}

\noindent
replacing $\ddot \lambda$ by (\ref{ddl12}) en (\ref{menho}), 
we obtain

\begin{equation}
- e^{(\nu-\lambda)/2} R = \left(\rho+P_r\right) \dot\omega + 
\dot Q e^{\lambda/2}
\label{pfR}
\end{equation}

\noindent
where $R$ denotes the left-hand side of the TOV equation, i.e.

\begin{eqnarray}
R & \equiv &  \frac{dP_r}{dr} + \frac{4\pi r P_r^2}{1-2m/r} + 
\frac{P_r m}{r^2 \left(1-2m/r\right)} + 
\frac{4\pi r \rho P_r}{1-2m/r} + \nonumber \\ 
 &  & + \frac{\rho m}{r^2 \left(1-2m/r\right)} - 
\frac{2\left(P_\bot - P_r\right)}{r} \nonumber \\
& = & P'_r + \frac{\nu'}{2} \left(\rho + P_r\right) -
\frac{2}{r} \left(P_\bot - P_r\right) 
\label{Rfr}
\end{eqnarray}

\noindent
The physical meaning of $R$ is clearly inferred from (\ref{Rfr}). 
It represents the total force (gravitational + pressure gradient + 
anisotropic term) acting on a given fluid element. Obviously, 
$R>0/R<0$ means that the total force is directed $inward/outward$ of 
the sphere.

\noindent
Let us now turn back to thermal conduction equation (\ref{Catrel}). 
Evaluating its $t$-component (given by Eq.(\ref{com0}))
immediately after perturbation, we obtain for the first initial 
configuration (static), an identity. Whereas the second case 
(slowly evolving) leads to

\begin{equation}
\omega \left(T' + T \frac{\nu'}{2}\right) = 0
\label{cs2}
\end{equation}

\noindent
which is to be expected, since before perturbation, in the 
slowly evolving regime, we have according to Eckart-Landau 
(valid in this regime)

\begin{equation}
Q = - \kappa e^{-\lambda} \left(T' + \frac{T \nu'}{2}\right)
\label{EL}
\end{equation}

\noindent
Therefore, the quantity in bracket is of order $Q$. Then 
immediately after perturbation this quantity is still of 
order $O(\omega)$, which implies (\ref{cs2}).

\noindent
The corresponding evolution of the $r$-component of the equation 
(\ref{Catrel}) yields, for the initially static configuration

\begin{equation}
\tau \dot Q e^{\lambda/2} = - \kappa T \dot\omega 
\label{Cat1}
\end{equation}

\noindent
where the fact has been used that after perturbation

\begin{equation}
Q = 0 \quad \Longrightarrow \quad T' = - \, \frac{T \nu'}{2}
\label{impl}
\end{equation}

\noindent
For the second case, the $r$-component of heat transport equation 
yields also (\ref{Cat1}), since after perturbation the value of $Q$ 
is still given by (\ref{EL}), up to $O(\omega)$ terms.

\noindent
Finally, combining (\ref{pfR}) and (\ref{Cat1}) we obtain

\begin{equation}
\dot\omega = - \frac{e^{(\nu-\lambda)/2} R}{\left(\rho+P_r\right)} 
\times 
\frac{1}{\left(1 - \frac{\kappa T}{\tau \left(\rho+P_r\right)}\right)}
\label{exmin}
\end{equation}

\noindent
or,

\begin{equation}
- e^{(\nu-\lambda)/2} R = 
\left(\rho + P_r\right) \dot \omega \left(1 - \alpha\right)
\label{Ralfa}
\end{equation}

 with  $\alpha$ defined  by 

\begin{equation}
\alpha \equiv \frac{\kappa T}{\tau \left(\rho + P_r\right)}
\label{alfa}
\end{equation}
\noindent
Let us first consider the $\alpha=0$ case. Then, (\ref{Ralfa})
has the obvious ``Newtonian'' form

\centerline{Force $=$ mass $\times$ acceleration}

\noindent
since, as it is well known, $\left(\rho + P_r\right)$ represents 
the inertial mass density and by ``acceleration'' we mean the time derivative 
of $\omega$. In this case ($\alpha=0$), an 
$outward/inward$ acceleration ($\dot \omega>0/\dot \omega<0$) is 
associated with an $outwardly/inwardly$ ($R<0/R>0$) directed total 
force (as one expects!). 

\noindent
However, in the general case ($\alpha \not = 0$) the situation 
becomes quite different. Indeed, the inertial mass term is now 
multiplied by ($1-\alpha$), so that if $\alpha=1$, we obtain that 
$\dot \omega \not = 0$ even though $R=0$. This decreasing  of the ``effective inertial mass density''( vanishing when $\alpha=1$), has been shown to occur in the general 
(non--spherical symmetric case)  (see \cite{eim} and references therein)

\noindent
Next, evaluating (\ref{W}) immediatly  after the system leaves the equilibrium (or quasi--equilibrium) and using (\ref{ddl12}) and (\ref{pfR}), (\ref{Ralfa}) may be written as 

\begin{equation}
- 3 e^{(\nu-\lambda)/2}\frac{\left[W-W_{(s)}\right]}{2 \pi r^4} = 
\left(\rho + P_r\right) \dot \omega \left(1 - \alpha\right)
\label{Ralfa2}
\end{equation}

Therefore, $\dot\omega \not=0$ implies $W \not=W_{(s)}$, and viceversa, unless either $(\rho+P_r)=0$ (in the non--dissipative case) or $\alpha=1$ (in the dissipative case),.

Next, assuming $\dot \omega=0$, it can be shown without difficulty, that
\begin{equation}
- 3 e^{(\nu-\lambda)/2}\frac{\frac {\partial^{n}}{\partial t^{n}}\left[W-W_{(s)}\right]}{2 \pi r^4} = 
\left(\rho + P_r\right) \frac {\partial^{n}\omega}{\partial t^{n}} \left(1 - \alpha\right)
\label{Ralfan}
\end{equation}
for any $n>1$.
Thus, unless either $(\rho+P_r)=0$ (in the non--dissipative case) or $\alpha=1$ (in the dissipative case), if the Weyl tensor does not change with respect to its value in the equilibrium (or quasi--equilibrium ) state,
 the system will not abandon such state.

\section{Conclusions}

We have seen that any departure from equilibrium (or quasi--equilibrium) of 
a spherically symmetric distribution of matter, is tightly controlled by changes in the Weyl tensor with respect to its value in equilibrium (or quasi--equilibrium),
as indicated by (\ref{Ralfa2}) and (\ref{Ralfan}). As an obvious consequence of this, it follows that a conformally flat fluid distribution in state of equilibrium 
(or quasi--equilibrium) will depart from such state, only if it ceases to be conformally flat (except for the two cases already mentioned).

\noindent
Finally, it is worth mentioning that the last term in (\ref{Catrel}) is frequently omitted (the so-called 
``truncated'' theory) \cite{TP}. In the context of this work both 
components of this term vanish and therefore all results found above 
are independent of the adopted theory (Israel-Stewart or truncated).

\section{Acknowledgments}
We acknowledge financial assistance under grant
BFM2000-1322 (M.C.T. Spain) and from C\'atedra-FONACIT, under grant
2001001789. .


\end{document}